\documentclass{PoS}

\newcommand{\beq}{\begin{equation}}
\newcommand{\enq}{\end{equation}}
\def\bx{{\bf x}}
\def\co{{\cal O}}
\def\svev#1{\left\langle #1\right\rangle}       % variable < >

\title{Spectroscopy of  SU(4) lattice gauge theory with fermions in the
  two index anti-symmetric representation}

\ShortTitle{SU(4) sextet spectrum}

\author{Thomas DeGrand$^1$, \speaker{Yuzhi Liu$^1$} and Ethan T.~Neil$^{1,2}$\\
        $^1$Department of Physics,
        University of Colorado, Boulder, CO 80309, USA\\
        $^2$RIKEN-BNL Research Center, Brookhaven National Laboratory, Upton, NY USA\\
        Email:  \email{degrand@pizero.colorado.edu}, \email{yuzhi.liu@colorado.edu}, 
                \email{ethan.neil@colorado.edu}}
\author{Yigal Shamir and Benjamin Svetitsky\\
        Raymond and Beverly Sackler School of Physics and Astronomy,
        Tel~Aviv University, 69978
        Tel~Aviv, Israel\\
        Email: \email{shamir@post.tau.ac.il}, \email{bqs@julian.tau.ac.il}}

\abstract{
We present a study of spectroscopy of $SU(4)$ lattice gauge theory coupled to 
two flavors of Dirac fermions in the anti-symmetric two index representation.
The fermion representation is real, 
and the pattern of chiral symmetry breaking is $SU(2N_f)\rightarrow SO(2N_f)$
with $N_f$ flavors of Dirac fermions.
It is an interesting generalization of QCD, for several reasons:
it allows direct exploration of an alternate large $N_c$ expansion,
it can be simulated at non-zero chemical potential with no sign problem,
and several UV completions of composite Higgs systems are built on it.
We present preliminary results on the baryon and meson spectra of the theory 
and compare them with $SU(3)$ results and with expectations for large $N_c$ scaling.
}   

\FullConference{The 32nd International Symposium on Lattice Field Theory,\\
		23-28 June, 2014\\
		Columbia University New York, NY}

\begin{document}
%%%%%%%%%%%%%%%%%%%%%%%%%%%%%%%%%%%%%%%%%%%%%%%%%%%%%%%%%%%%%%%%%%%%%%
\section{Introduction}

The authors of this paper are involved in a diverse set of
projects involving $SU(4)$ gauge theory with various numbers of 
flavors of degenerate mass fermions in the two-index antisymmetric (AS2) 
representation of the gauge group, which is a sextet for $SU(4)$. 
These systems are interesting for a variety of reasons:

First, they are confining and chirally broken systems with similarities to
ordinary $N_c=3$ QCD. In fact, there is an alternate large-$N_c$ limit of ordinary QCD 
in which the fermions live in an AS2 representation. For $N_c=3$, AS2 quarks inhabit the 
$\bar 3$ representation. The story goes back to  \cite{Corrigan:1979xf}.
It reappears in more modern guises in,
for example, \cite{Armoni:2003fb,Cherman:2012eg}.
Lattice simulation can test the expected large-$N_c$ regularities, as it has for the
usual 't Hooft limit of fixed $N_f$ fundamental representation fermions at varying $N_c$.
(An assortment of recent results includes  \cite{Lucini:2012gg,Bali:2013kia}.)

Next, they form a chirally broken system with some differences compared to
ordinary $N_c=3$ QCD. Because the fermions are in a  real representation of the gauge group,
the pattern of chiral symmetry breaking is not $SU(N_f)\otimes SU(N_f)\rightarrow
SU(N_f)$; it is $SU(2N_f)\rightarrow SO(2N_f)$ (all for $N_f$ flavors of Dirac fermions)
\cite{Peskin:1980gc}. The reality of the representation allows quarks and antiquarks to 
mix under global flavor rotations. 
In particular, the $N_f=2$ theory has nine Goldstone bosons,
which may be classified as isospin $I=1$ triplets of $q \bar q$, 
 $q  q$, and  $\bar q  \bar q$.

Third, reality of the representation means that finite density simulations
do not suffer from a sign problem. This is similar to the situation for $N_c=2$
with fundamental representation fermions \cite{Kogut:2000ek}.
There is a literature of predictions for $SU(4)$ \cite{Blake:2012dp}, which we can explore.

Finally, members of this family play a role in composite Higgs studies. For example,
the Littlest Higgs model \cite{ArkaniHamed:2002qy} relies on the non-linear sigma model 
$SU(5)/SO(5)$. Examples of proposed $SU(4)$ UV completions of composite Higgs models, 
mostly involving 5 Majorana fermions, are given in Refs.~\cite{Vecchi:2013bja}.

In this note we describe results relevant to the first of these points.
The details of the calculations will be presented in our longer paper \cite{su4_2014}.

%%%%%%%%%%%%%%%%%%%%%%%%%%%%%%%%%%%%%%%%%%%%%%%%%%%%%%%%%%%%%%%%%%%%%%
\section{Lattice setup and observables}
%%%%%%%%%%%%%%%%%%%%%%%%%%%%%%%%%%%%%%%%%%%%%%%%%%%%%%%%%%%%%%%%%%%%%%

We use the usual Wilson plaquette gauge action and Wilson-clover fermions with 
nHYP smeared links as the gauge connections.
The bare gauge coupling $g$ is defined through $\beta=2N_c/g^2$.
The bare quark mass $m$ is introduced through the hopping parameter $\kappa$. 
The clover coefficient is fixed at its tree level value, $c_{sw}=1$.

Simulations were done at four different $\kappa$ values at a bare coupling $\beta=9.6$. 
The lattice volume is fixed to be $16^3\times32$. In addition, we calculated spectroscopy at
four more partially quenched (PQ) points based on one dynamical data set. 

Our large-$N_c$ comparisons are done against simulations of $SU(3)$ gauge theory 
with $N_f=2$ fundamental fermions. Five different $\kappa$ values were used  
at one fixed gauge coupling. 
Previously generated quenched $SU(N_c)$ theories, with $N_c=3$, 5, and~7 
are also used for comparison \cite{DeGrand:2012hd}. All these data sets
 had the same volume, $16^3\times 32$.
For comparison among different theories, we fix the lattice spacings using $r_1$,
the shorter version \cite{Bernard:2000gd} of the Sommer~\cite{Sommer:1993ce} parameter, 
defined in terms of the force $F(r)$ between static quarks:
$r^2 F(r)=-1.0$ at $r=r_1$.

The pseudoscalar and vector meson decay constants $f_{PS}$ and $f_V$ are defined below
in Eqs.~(\ref{eq:fpi}) and (\ref{eq:fv});
 and the quark mass $m_q$ is defined from the axial Ward identity (AWI) Eq.~(\ref{eq:AWI}).
\beq
    \langle 0| \bar u \gamma_0 \gamma_5 d |PS\rangle = m_{PS} f_{PS},
    \label{eq:fpi}
\enq
\beq
    \langle 0| \bar u \gamma_i d  | V\rangle = m_V^2 f_V \epsilon_i ,
    \label{eq:fv}
\enq
\beq
    \partial_t \sum_\bx \svev{A_0^a(\bx,t)\co^a} = 2m_q \sum_\bx \svev{ P^a(\bx,t)\co^a}.
    \label{eq:AWI}
\enq
$A_\mu^a=\bar \psi \gamma_\mu\gamma_5 (\tau^a/2)\psi$ is the axial current,
$\vec\epsilon$ is a polarization vector,
 $P^a=\bar \psi \gamma_5 (\tau^a/2)\psi$ is the pseudoscalar density, and
$\co^a$ is a  source.
In our normalization convention $f_{PS}\approx 132$ MeV. 
In Eqs.~(\ref{eq:fpi}) and (\ref{eq:fv}), the lattice decay constants need to be renormalized 
by a field rescaling and the corresponding $Z$ factors to get the continuum quantities.
For the pseudoscalar decay constant, we have
\beq
    f_{PS}^{cont} = \left(1 - \frac34\frac{\kappa}{\kappa_c}\right) Z_{PS} f_{PS}^{latt}.
    \label{eq:fpsv}
\enq
There is a similar equation for the vector case.
For our case, the $Z$ factors are close to unity \cite{su4_2014, DeGrand:2002vu}.

%%%%%%%%%%%%%%%%%%%%%%%%%%%%%%%%%%%%%%%%%%%%%%%%%%%%%%%%%%%%%%%%%%%%%%
\section{Phase diagram}
%%%%%%%%%%%%%%%%%%%%%%%%%%%%%%%%%%%%%%%%%%%%%%%%%%%%%%%%%%%%%%%%%%%%%%

\begin{figure}
\begin{center}
\includegraphics[width=0.4\textwidth,clip]{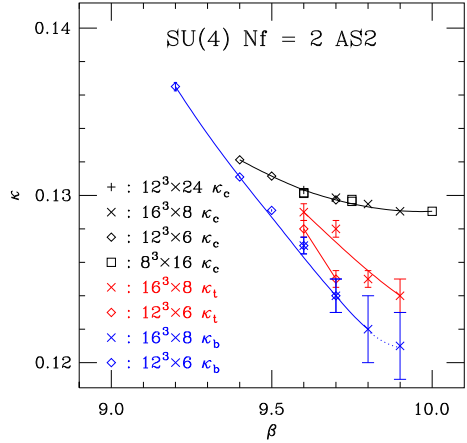}
\end{center}
\caption{Phase diagram of the $SU(4)$ AS2 theory in the ($\beta$, $\kappa$) plane. 
The solid lines are drawn to guide the eye and are not a fit to the data. 
From right to left: the $\kappa_c$ line, the thermal transition lines $\kappa_t(N_t = 8)$ 
and $\kappa_t(N_t = 6)$, and the bulk transition line $\kappa_b$.
The dotted line indicates weakening of the bulk transition to a crossover.
\label{fig:phase_diagram}}
\end{figure}

Before computing spectroscopy, we have to map out the phase structure of the 
system in the $(\beta, \kappa)$ plane. The result is shown in Fig.~\ref{fig:phase_diagram}. 
It is a bit complicated. Here are the ingredients:

Running along the top right side of the figure is the $\kappa_c$ line, 
where the AWI quark mass vanishes.
The steeply falling line on the left is a bulk transition. It appears to be first order
out to $\beta=9.7$, and then seems to turn into a crossover. When it is first order,
the quark mass jumps discontinuously. We believe that the $\kappa_c$ line disappears
when it encounters this transition, so that at sufficiently strong coupling there is 
no zero quark mass point for this lattice action.

The region between the two lines contains the desired confining and chirally broken phase.
We did simulations on asymmetric lattices and observed finite temperature transitions 
from a confined to a deconfined phase. These lines move to bigger $\beta$ as
the temporal size of the lattice increases.
We wanted to simulate at lattice spacings which were neither too large or too small.
We settled on a line varying $\kappa$ at $\beta=9.6$.

%%%%%%%%%%%%%%%%%%%%%%%%%%%%%%%%%%%%%%%%%%%%%%%%%%%%%%%%%%%%%%%%%%%%
%%%%%%%%%%%%%%%%%%%%%%%%%%%%%%%%%%%%%%%%%%%%%%%%%%%%%%%%%%%%%%%%%%%%

%%%%%%%%%%%%%%%%%%%%%%%%%%%%%%%%%%%%%%%%%%%%%%%%%%%%%%%%%%%%%%%%%%%%%%
\section{Large $N_c$ scaling tests}
%%%%%%%%%%%%%%%%%%%%%%%%%%%%%%%%%%%%%%%%%%%%%%%%%%%%%%%%%%%%%%%%%%%%%%

We computed the masses of the pseudoscalar and vector mesons and their decay constants, 
and the masses of $J=0$ and $J=1$ diquark states, which were degenerate
with their mesonic analogs, as expected (this should no longer be true if a chemical potential is 
turned on).
Meson masses are expected to be $N_c$ independent, regardless of the fermion representation.

Pseudoscalar and vector meson decay constants $f$ scale differently with $N_c$ in the fundamental and 
AS2 representations. The expected large $N_c$ scaling behavior is
\beq
  f \sim \left\{
  \begin{array}{r l}
    \sqrt{N_c} & \quad \textrm{fundamental,}\\
    N_c & \quad \textrm{AS2.}
  \end{array} 
  \right.\
\label{eq:fpi_fv_scaling}
\enq

In leading order in $N_c$, baryon masses scale with the number of quarks ($N_b$) in the baryon, 
with corrections. 
$N_b=N_c$ for  fundamental representation fermions, of course, and $N_b=N_c(N_c-1)/2$ 
for  AS2 fermions \cite{Bolognesi:2006ws}. This means that  $N_b=6$ for our $N_c=4$ case. 
This is easy to understand by noting that the AS2 representation of $SU(4)$ is 
equivalent to the vector representation of $SO(6)$, and the color singlet baryon wave 
function is just the antisymmetric product of six vectors. 
At order $1/N_b$, the baryon mass $M_B$ is given by  the rotor formula~\cite{Adkins:1983ya,Jenkins:1993zu}
\beq
    M_B(J)~\approx~N_b m_0 + B\frac{J(J+1)}{N_b}.
    \label{eq:jsplit}
\enq
The parameters $m_0$ and $B$ depend on the quark mass.
These are just the leading terms in a $1/N_c$ expansion. For example, 
$m_0 = m_{00} + (1/N_c) m_{01}+ (1/N_c^2) m_{02} + \cdots$. 
The terms in the expansion, such as $m_{01}$, are expected to have some ``typical hadronic size.'' 
This generic behavior is also expected for meson masses and decay constants.
%%%%%%%%%%%%%%%%%%%%%%%%%%%%%%%%%%%%%%%%%%%%%%%%%%%%%%%%%%%%%%%%%%%%%
\begin{figure}
\begin{center}
\includegraphics[width=0.8\textwidth,clip]{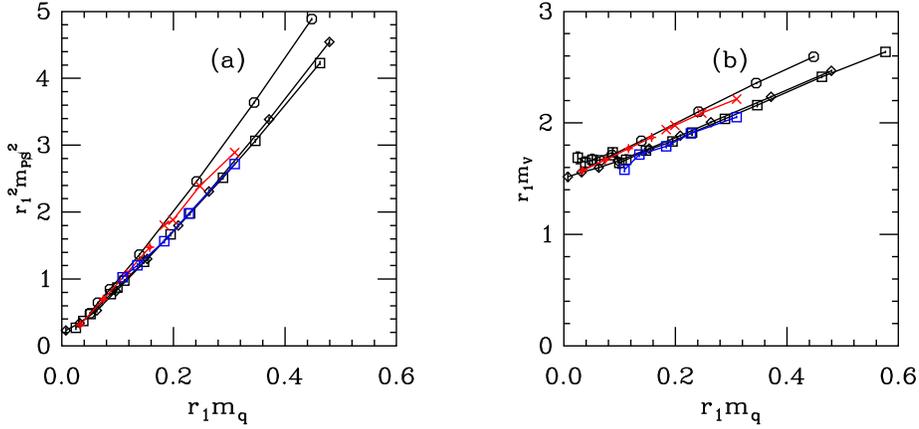}
\end{center}
\caption{Meson spectroscopy. On the left, the squared pseudoscalar mass scaled by  $r_1^2$, on the right,
$r_1$ times the vector meson mass. The abscissa is $r_1$ times the AWI quark mass.
The data sets are:
black squares for quenched $SU(3)$ fundamentals, black diamonds for quenched $SU(5)$ fundamentals,
black octagons for quenched $SU(7)$ fundamentals,
red crosses for $SU(4)$ AS2; the fancy diamonds are the PQ data. Finally, the blue squares
are  $SU(3)$ with two dynamical, fundamental flavors.
\label{fig:mesons}}
\end{figure}
%%%%%%%%%%%%%%%%%%%%%%%%%%%%%%%%%%%%%%%%%%%%%%%%%%%%%%%%%%%%%%%%%%%%%

In Fig.~\ref{fig:mesons}, we plot the data for the pseudoscalar and vector meson masses as a 
function of the AWI quark mass $m_q$. 
The weak dependence of meson masses on $N_c$ and representation confirms large-$N_c$ expectations.

\begin{figure}
\begin{center}
\includegraphics[width=0.8\textwidth,clip]{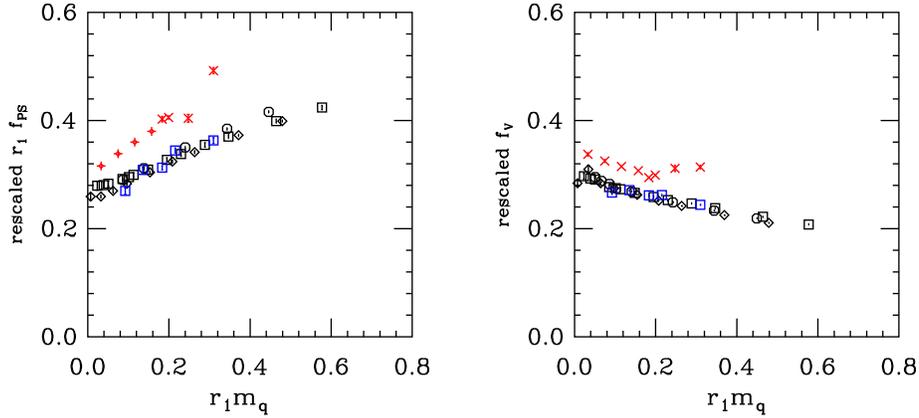}
\caption{Pseudoscalar and vector meson decay constants. 
The abscissa is $r_1$ times the AWI quark mass.
The meaning of the symbols is the same as in Fig.~{\protect\ref{fig:mesons}}.
The data are rescaled according to Eq.~({\protect\ref{eq:fpi_fv_scaling}}) as described in the text. 
\label{fig:fpi_fv_mq}}
\end{center}
\end{figure}

To compare decay constants at different $N_c$, we follow Eq.~(\ref{eq:fpi_fv_scaling}) and 
rescale the fundamental representation data by $\sqrt{3/N_c}$, and the AS2 data by $3/N_c$.
In Fig.~\ref{fig:fpi_fv_mq}, we plot the rescaled pseudoscalar and vector meson decay constants 
in $r_1$ units. 
Dynamical $SU(3)$ data overlap well with all the different $N_c$ quenched fundamental ones.
The $SU(4)$ AS2 data is consistently above the fundamental ones, but the discrepancy is less than $20\%$.
%%%%%%%%%%%%%%%%%%%%%%%%%%%%%%%%%%%%%%%%%%%%%%%%%%%%%%%%%%%%%%%%%%%%%
%%%%%%%%%%%%%%%%%%%%%%%%%%%%%%%%%%%%%%%%%%%%%%%%%%%%%%%%%%%%%%%%%%%%%

\begin{figure}
\begin{center}
\includegraphics[width=0.45\textwidth]{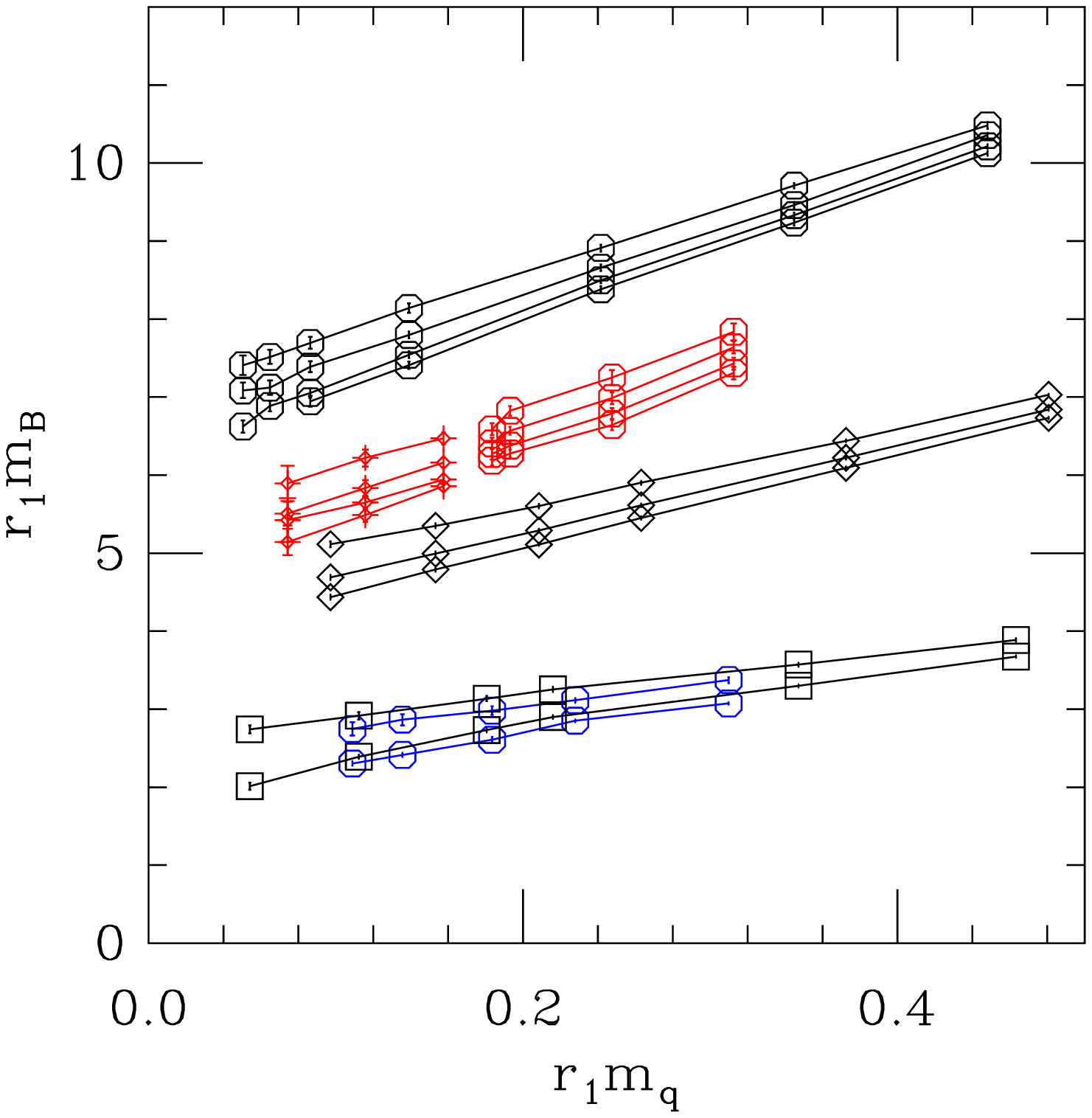}
\end{center}
\caption{Baryons. The black data are (from the top) quenched $SU(7)$, $SU(5)$ and $SU(3)$ data.
The blue octagons are $SU(3)$ with dynamical fermions. 
The red lines are the six quark baryons in $SU(4)$ AS2, 
octagons for dynamical and fancy diamonds for partially quenched. 
Higher $J$ states lie higher in mass and equal $J$ value points are connected by lines.
\label{fig:baryon}}
\end{figure}

\begin{figure}
\begin{center}
\includegraphics[width=0.8\textwidth,clip]{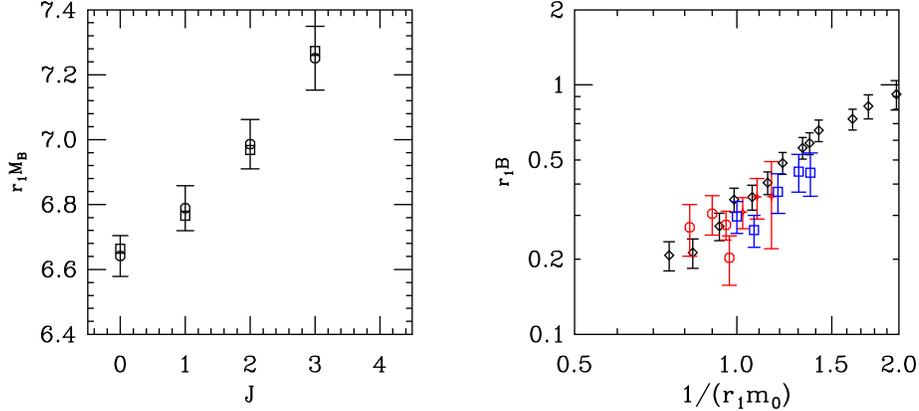}
\end{center}
\caption{Left: Fit to rotor formula. $SU(4)$ AS2; $\kappa=0.1285$.
Octagons with error bars are the data points; 
squares the best fit values.
Right: $B$ vs. $m_0$ from the rotor formula; 
black diamonds from quenched $SU(3)$, 
blue squares from full $SU(3)$.
The $SU(4)$ data are shown as red octagons for the dynamical sets and 
fancy diamonds for the partially quenched set.
\label{fig:f01285}}
\end{figure}

Baryon mass data are shown in Fig.~\ref{fig:baryon}. 
%In all cases higher $J$ states lie higher in mass. The lines connect them.
%%%%%%%%%%%%%%%%%%%%%%%%%%%%%%%%%%%%%%%%%%%%%%%%%%%%%%%%%%%%%%%%%%%%%
%%%%%%%%%%%%%%%%%%%%%%%%%%%%%%%%%%%%%%%%%%%%%%%%%%%%%%%%%%%%%%%%%%%%%
To compare to the rotor formula, we fit the data with Eq.~(\ref{eq:jsplit}) treating $m_0$ and $B$ 
as free parameters.
The fit results are shown in the left panel of Fig.~\ref{fig:f01285} for AS2 data at one quark mass, 
corresponding to $\kappa=0.1285$.
The squares are the fit results and the octagons with error bars are the data points.
The correlation between the parameters $m_0$ and $B$ at different quark masses is shown in
the right panel of Fig.~\ref{fig:f01285}.
The slope of $r_1B$ versus $1/({r_1}{m_0})$ is around one in the log-log plot. 
This suggests that the parameter $B$ is inversely proportional to $m_0$: 
this is consistent with the rotor formula, since $N_B/(2B)$ is the baryon's moment of inertia.
%%%%%%%%%%%%%%%%%%%%%%%%%%%%%%%%%%%%%%%%%%%%%%%%%%%%%%%%%%%%%%%%%%%%%
%%%%%%%%%%%%%%%%%%%%%%%%%%%%%%%%%%%%%%%%%%%%%%%%%%%%%%%%%%%%%%%%%%%%%

%%%%%%%%%%%%%%%%%%%%%%%%%%%%%%%%%%%%%%%%%%%%%%%%%%%%%%%%%%%%%%%%%%%%%%
\section{Conclusions}
%%%%%%%%%%%%%%%%%%%%%%%%%%%%%%%%%%%%%%%%%%%%%%%%%%%%%%%%%%%%%%%%%%%%%%
Large-$N_c$ scaling certainly seems to describe all of our data, both with fundamental and AS2 fermions. 
Even the quantities with the poorest agreement, the decay constants, show only a twenty per cent discrepancy. 
Phenomenologists commonly use large-$N_c$ scaling to move from QCD to other confining theories. 
Lattice simulations show that this is a reasonable thing to do.

Our large-$N_c$ story for AS2 fermions is still incomplete. With only two $N_c$'s, one cannot 
do any kind of detailed analysis. 
Additionally, $SU(4)$ with AS2 fermions is also special in its pattern of chiral symmetry breaking 
compared to all other $N_c$'s.  We plan to continue our studies of this curious system. 

%%%%%%%%%%%%%%%%%%%%%%%%%%%%%%%%%%%%%%%%%%%%%%%%%%%%%%%%%%%%%%%%%%%%%%

\begin{acknowledgments}
T.~D. would like to thank Richard Lebed for correspondence and conversations.
B.~S. thanks the University of Colorado for hospitality, as well as the Yukawa 
Institute for Theoretical Physics at Kyoto University.
This work was supported in part by the U.~S. Department of Energy and by the 
Israel Science Foundation under Grant no.~449/13.
Computations were performed using USQCD resources at Fermilab and
on the University of Colorado theory group's cluster.
Our computer code is based on version 7 of the publicly available code of the 
MILC collaboration.
\end{acknowledgments}

\end{document}